# On the Linear Precoder Design for MIMO Channels with Finite-Alphabet Inputs and Statistical CSI

Weiliang Zeng*, Chengshan Xiao†, Mingxi Wang†, and Jianhua Lu*
*State Key Laboratory on Microwave and Digital Communications
Tsinghua National Laboratory for Information Science and Technology
Dept. of Electronic Engineering, Tsinghua University, Beijing, China
†Dept. of Electrical & Computer Engineering, Missouri University of Science and Technology, MO, USA

***Abstract*—This paper investigates the linear precoder design that maximizes the average mutual information of multiple-input multiple-output channels with finite-alphabet inputs and statistical channel state information known at the transmitter. This linear precoder design is an important open problem and is extremely difficult to solve: First, average mutual information lacks closed-form expression and involves complicated computations; Second, the optimization problem over precoder is nonconcave. This study explores the solution to this problem and provides the following contributions: 1) A closed-form lower bound of average mutual information is derived. It achieves asymptotic optimality at low and high signal-to-noise ratio regions and, with a constant shift, offers an accurate approximation to the average mutual information; 2) The optimal structure of the precoder is revealed, and a unified two-step iterative algorithm is proposed to solve this problem. Numerical examples show the convergence and the efficacy of the proposed algorithm. Compared to its conventional counterparts, the proposed linear precoding method provides a significant performance gain.**

## I. Introduction

The theoretic limit on the information rate that a communication channel can support with arbitrary low probability of error is referred to as *channel capacity*. This capacity is achievable with independent Gaussian distributed inputs for parallel additive white Gaussian noise (AWGN) channels and with correlated Gaussian inputs for multiple-input multiple-output (MIMO) channels [1].

Even though Gaussian inputs are theoretically optimal, they are rarely realized in practice. Alternatively, inputs are usually taken from a finite-alphabet constellation set, such as phase shift keying (PSK) modulation and quadrature amplitude modulation (QAM), which departs significantly from the Gaussian assumption. Therefore, there can be a big performance gap between the precoding schemes designed from the standpoint of finite-alphabet inputs and those designed with Gaussian-input assumptions. For example, in [2], the optimal power allocation for parallel Gaussian channels with finite-alphabet inputs is obtained; for the case of MIMO channels, the necessary condition satisfied by the optimal precoder is given in [3]. Optimization of the precoder using gradient-descent method is introduced in [4]; Optimization utilizing the structure of the optimal precoder is considered in [5]–[7].

The above-stated results and precoding algorithms hold when the transmitter is able to accurately track the instantaneous channel state information (CSI). For fast fading channels, long-term channel statistics is more plausible to exploit because it varies with the antenna parameters and the surrounding environment and thus may change very slowly.

This study explores the linear precoder that maximizes the average mutual information with statistical CSI. It starts by decomposing the precoder, by the singular value decomposition (SVD), into three components: left singular vectors, diagonal power allocation matrix, and right singular vectors, and proves the left singular vectors equal the eigenvectors of the transmit correlation matrix. Due to the prohibitive complexity of evaluating average mutual information, a closed-form lower bound is derived. Interestingly, with a constant shift, the lower bound function offers a very good approximation to the average mutual information, and using the bound for precoder design achieves the optimality asymptotically in the low and high signal-to-noise ratio (SNR) regions. Therefore, this paper proposes to use the bound as an alternative and develops an iterative algorithm, based on convex optimization and optimization on the Stiefel manifold, to obtain good solutions to power allocation matrix and right singular vectors.

*Notation:* Boldface uppercase letters denote matrices, boldface lowercase letters denote column vectors, and italics denote scalars. The superscripts $(\cdot)^T$ and $(\cdot)^H$ stand for transpose and Hermitian operations, respectively; $[\mathbf{A}]_{i,j}$ denotes the $(i,j)$-th element of matrix $\mathbf{A}$; $\mathsf{Tr}(\mathbf{A})$ denotes the trace operation; $\mathbf{I}$ represent an identity matrix. $\mathsf{E}$ denotes statistical expectation, and $\mathbb{C}$ denotes the complex spaces. All logarithms are base 2.

## II. System Model and Preliminaries

Consider a MIMO system over frequency flat fading with $N_t$ transmit antennas and $N_r$ receive antennas. Let $\mathbf{x} \in \mathbb{C}^{N_t}$ be a transmit signal vector with zero mean and unit covariance; the receive signal $\mathbf{y} \in \mathbb{C}^{N_r}$ is given by

$$\mathbf{y} = \mathbf{HPx} + \mathbf{n} \tag{1}$$

where $\mathbf{H} \in \mathbb{C}^{N_r \times N_t}$ is a random channel matrix whose $(i,j)$-th entry is the complex propagation coefficient between the $j$-th transmit antenna and the $i$-th receive antenna; $\mathbf{P} \in \mathbb{C}^{N_t \times N_t}$ is a linear precoding matrix; $\mathbf{n} \in \mathbb{C}^{N_r}$ is an independent and identically distributed (i.i.d.) zero-mean circularly-symmetric Gaussian noise with covariance $\sigma^2\mathbf{I}$.

For doubly correlated MIMO channels, the channel matrix $\mathbf{H}$ can be modeled as

$$\mathbf{H} = \mathbf{\Psi_r}^{1/2}\mathbf{H_w}\mathbf{\Psi_t}^{1/2} \tag{2}$$

This work was supported in part by the National Basic Research Program of China (2007CB310601) and the U.S. National Science Foundation under Grant CCF-0915846. It has been carried out while W. Zeng is a visiting scholar at Missouri University of Science and Technology.

where $\mathbf{H_w} \in \mathbb{C}^{N_r \times N_t}$ is a complex matrix with i.i.d. zero-mean and unit variance Gaussian entries; $\mathbf{\Psi_t} \in \mathbb{C}^{N_t \times N_t} > 0$ and $\mathbf{\Psi_r} \in \mathbb{C}^{N_r \times N_r} > 0$, respectively, are transmit and receive correlation matrices known by transmitter.

With the product of $\mathbf{H}$ and $\mathbf{P}$ known at the receiver and input signal drawn from the $M$-ary equiprobable finite-alphabet constellation, the average mutual information between $\mathbf{x}$ and $\mathbf{y}$, $\mathcal{I}(\mathbf{x};\mathbf{y})$, is given by [3]:

$$\mathcal{I}(\mathbf{x};\mathbf{y}) = \mathsf{E}_{\mathbf{H_w}} \hat{\mathcal{I}}(\mathbf{x};\mathbf{y}) \tag{3}$$

in which $\hat{\mathcal{I}}(\mathbf{x};\mathbf{y})$ is the instantaneous mutual information:

$$\hat{\mathcal{I}}(\mathbf{x};\mathbf{y}) = N_t \log M - \frac{1}{M^{N_t}} \sum_{m=1}^{M^{N_t}} \mathsf{E}_{\mathbf{n}} \log \sum_{k=1}^{M^{N_t}} \exp\left(-d_{m,k}\right),$$

where $d_{m,k} = (\|\mathbf{H}\mathbf{P}\mathbf{e}_{mk} + \mathbf{n}\|^2 - \|\mathbf{n}\|^2)/\sigma^2$, $\|\cdot\|$ denotes the Euclidean norm of a vector, and $\mathbf{e}_{mk} = \mathbf{x}_m - \mathbf{x}_k$. Both $\mathbf{x}_m$ and $\mathbf{x}_k$ contain $N_t$ symbols, taken from signal constellation.

Considering the unitarily invariant of Euclidean norm, the following relationship can be identified:

$$\mathcal{I}(\mathbf{x};\mathbf{y}) = \mathcal{I}(\mathbf{x};\mathbf{U}\mathbf{y}) \quad \text{and} \quad \mathcal{I}(\mathbf{x};\mathbf{y}) \neq \mathcal{I}(\mathbf{U}\mathbf{x};\mathbf{y}) \tag{4}$$

which implies that linear precoder, even a unitary one, may change the average mutual information of MIMO systems.

The objective of this work is to develop efficient algorithm to find a linear precoding solution that maximizes the average mutual information in (3). The optimization is carried out over all possible $N_t \times N_t$ complex precoding matrices with transmit power constraint:

$$\begin{aligned} &\text{maximize} && \mathcal{I}(\mathbf{x};\mathbf{y}) \\ &\text{subject to} && \mathsf{Tr}(\mathbf{P}\mathbf{P}^H) \leq N_t. \end{aligned} \tag{5}$$

Since $\mathcal{I}(\mathbf{x};\mathbf{y})$ is an increasing function of SNR (*i.e.*, $1/\sigma^2$), the optimal precoder should use the maximum available power; that is, the inequality constraint can be replaced by the equality constraint: $\mathsf{Tr}(\mathbf{P}\mathbf{P}^H) = N_t$.

The obstacles in the way to solve problem (5) are twofold. First, the closed-form objective function is lacking (see [4] for the case of instantaneous CSI); second, the optimization problem is nonconcave and extremely difficult even for some specific cases (see [5] for the case of instantaneous CSI with real-valued channels). The next section will explore the structure of the optimal precoding matrix and will provide a closed-form lower bound to the objective function.

## III. Optimal Precoding Structure and Average Mutual Information Bound

From eigenvalue decomposition, the correlation matrices $\mathbf{\Psi_t}$ and $\mathbf{\Psi_r}$ can be expressed as

$$\mathbf{\Psi_t} = \mathbf{U_t}\mathbf{\Sigma_t}\mathbf{U_t}^H \quad \text{and} \quad \mathbf{\Psi_r} = \mathbf{U_r}\mathbf{\Sigma_r}\mathbf{U_r}^H \tag{6}$$

where $\mathbf{U_t}$ and $\mathbf{U_r}$ are unitary matrices whose columns are eigenvectors of $\mathbf{\Psi_t}$ and $\mathbf{\Psi_r}$, and $\mathbf{\Sigma_t}$ and $\mathbf{\Sigma_r}$ are diagonal matrices whose diagonal entries are the eigenvalues of $\mathbf{\Psi_t}$ and $\mathbf{\Psi_r}$, respectively. Applying the property in (4) and the fact that random matrices $\mathbf{H_w}$ and $\tilde{\mathbf{H}}_\mathbf{w} = \mathbf{U_r}^H\mathbf{H_w}\mathbf{U_t}$ have the same statistics, the channel model (1) can be reduced to

$$\tilde{\mathbf{y}} = \tilde{\mathbf{H}}\mathbf{P}\mathbf{x} + \tilde{\mathbf{n}} \tag{7}$$

where $\tilde{\mathbf{H}} = \mathbf{\Sigma_r}^{\frac{1}{2}}\tilde{\mathbf{H}}_\mathbf{w}\mathbf{\Sigma_t}^{\frac{1}{2}}\mathbf{U_t}^H$, and $\tilde{\mathbf{y}}$ and $\tilde{\mathbf{n}}$ are the results when unitary transform $\mathbf{U_r}^H$ is applied on $\mathbf{y}$ and $\mathbf{n}$, respectively. Because maximizing $\mathcal{I}(\mathbf{x};\tilde{\mathbf{y}})$ based on the model (7) is equivalent to maximizing $\mathcal{I}(\mathbf{x};\mathbf{y})$ from the model (1), the sequel discussion is based on this simplified model.

The instantaneous mutual information $\hat{\mathcal{I}}(\mathbf{x};\tilde{\mathbf{y}})$ depends on $\mathbf{P}$ through $\mathbf{M} = \mathbf{P}^H\tilde{\mathbf{H}}^H\tilde{\mathbf{H}}\mathbf{P}$ (see [5], [6] for the case of real-valued channels and [7] for complex-valued channels), which is a function of the random matrix $\tilde{\mathbf{H}}_\mathbf{w}$. The expectation taken for average over $\tilde{\mathbf{H}}_\mathbf{w}$ thus equals the expectation over $\mathbf{M}$:

$$\mathcal{I}(\mathbf{x};\tilde{\mathbf{y}}) = \mathsf{E}_{\tilde{\mathbf{H}}_\mathbf{w}}\hat{\mathcal{I}}(\mathbf{M}) = \mathsf{E}_\mathbf{M}\hat{\mathcal{I}}(\mathbf{M}) \tag{8}$$

where $\hat{\mathcal{I}}(\mathbf{M})$ emphasizes the dependence of instantaneous mutual information on $\mathbf{M}$. Since $\mathcal{I}(\mathbf{x};\tilde{\mathbf{y}})$ is also a function of the random matrix $\mathbf{M}$, the value of $\mathcal{I}(\mathbf{x};\tilde{\mathbf{y}})$ changes based on its probability density function (PDF) [8]:

$$\begin{aligned} p(\mathbf{M}) =& \frac{1}{\tilde{\Gamma}_{N_r}(N_t)} \mathsf{det}(\mathbf{W})^{-N_t} \mathsf{det}(\mathbf{\Sigma_r})^{-N_r} \mathsf{det}(\mathbf{M})^{N_t - N_r} \\ & \times {}_0\tilde{F}_0^{(N_t)}\left(-\mathbf{W}^{-1}\mathbf{M}, \mathbf{\Sigma_r}^{-1}\right), \quad \mathbf{M} > 0 \end{aligned} \tag{9}$$

where $\tilde{\Gamma}_{N_r}(N_t)$ is related to $N_r$ and $N_t$; $\mathbf{W}$ is $\mathbf{P}^H\mathbf{U_t}\mathbf{\Sigma_t}\mathbf{U_t}^H\mathbf{P}$; ${}_0\tilde{F}_0^{(N_t)}(\cdot)$ is the hypergeometric function of two Hermitian matrices.

From (9), the distribution of $\mathbf{M}$ is determined by constant parameters $\mathbf{\Sigma_r}$ and $\mathbf{W}$. Consider SVD of the precoding matrix $\mathbf{P} = \mathbf{U_P}\mathbf{\Sigma_P}\mathbf{V_P}^H$, where $\mathbf{U_P}$ and $\mathbf{V_P}$ are unitary matrices, and $\mathbf{\Sigma_P}$ contains nonnegative diagonal entries.

*Proposition 1:* Given parameter $\mathbf{W} = \mathbf{P}^H\mathbf{U_t}\mathbf{\Sigma_t}\mathbf{U_t}^H\mathbf{P}$ of the distribution of random matrix $\mathbf{M}$, the precoder in the form $\mathbf{P} = \mathbf{U_t}\mathbf{\Sigma_P}\mathbf{V_P}^H$ minimizes the transmit power $\mathsf{Tr}(\mathbf{P}\mathbf{P}^H)$.

*Proof:* See Appendix 3.B in [9]. □

This result provides the design for the left singular vectors, which equal the eigenvectors of transmit correlation matrix $\mathbf{\Psi_t}$; it simplifies the channel model (7) to

$$\tilde{\mathbf{y}} = \mathbf{\Sigma_r}^{\frac{1}{2}}\tilde{\mathbf{H}}_\mathbf{w}\mathbf{\Sigma_t}^{\frac{1}{2}}\tilde{\mathbf{P}}\mathbf{x} + \tilde{\mathbf{n}} \tag{10}$$

and reduces the average mutual information function to

$$\begin{aligned} \mathcal{I}(\mathbf{x};\tilde{\mathbf{y}}) =& N_t \log M - N_r/\ln 2 - 1/M^{N_t} \\ & \times \sum_{m=1}^{M^{N_t}} \mathsf{E}_{\tilde{\mathbf{H}}_\mathbf{w}} \mathsf{E}_\mathbf{n} \log \sum_{k=1}^{M^{N_t}} \exp\left(-\frac{\|\mathbf{\Sigma_r}^{\frac{1}{2}}\tilde{\mathbf{H}}_\mathbf{w}\mathbf{\Sigma_t}^{\frac{1}{2}}\tilde{\mathbf{P}}\mathbf{e}_{mk} + \mathbf{n}\|^2}{\sigma^2}\right) \end{aligned}$$

where $\tilde{\mathbf{P}} = \mathbf{\Sigma_P}\mathbf{V_P}^H$ is the remaining part of $\mathbf{P}$. However, it is still difficult, if not impossible, to evaluate the multiple integral numerically (for $N_r \times N_t$ MIMO channels, $2(N_rN_t + N_r)$ integrals from $-\infty$ to $\infty$ need to be considered).

*Proposition 2:* The average mutual information of doubly correlated MIMO channels with finite-alphabet inputs can be lower bounded by

$$\begin{aligned} \mathcal{I}_L =& N_t \log M - N_r\left(1/\ln 2 - 1\right) - 1/M^{N_t} \\ & \times \sum_{m=1}^{M^{N_t}} \log \sum_{k=1}^{M^{N_t}} \prod_q \left(1 + \frac{r_q}{2\sigma^2}\mathbf{e}_{mk}^H\tilde{\mathbf{P}}^H\mathbf{\Sigma_t}\tilde{\mathbf{P}}\mathbf{e}_{mk}\right)^{-1} \end{aligned}$$

where $r_q$ denotes the $q$-th diagonal element of $\mathbf{\Sigma_r}$.

*Proof:* This bound can be proved by using Jensen's inequality directly. Details are omitted here for brevity. □

## IV. Precoder Design for Maximizing the Average Mutual Information

This section starts by proving the asymptotic optimality of maximizing the lower bound and then develops a unified algorithm to do that. Based on Proposition 2, the problem of maximizing $\mathcal{I}_L$ is equivalent to the following problem:

$$\begin{aligned} \text{minimize} \quad & \sum_{m=1}^{M^{N_t}} \log \sum_{k=1}^{M^{N_t}} \prod_{q} \left(1 + \frac{r_q}{2\sigma^2} \mathbf{e}_{mk}^H \tilde{\mathbf{P}}^H \mathbf{\Sigma_t} \tilde{\mathbf{P}} \mathbf{e}_{mk}\right)^{-1} \\ \text{subject to} \quad & \mathsf{Tr}(\tilde{\mathbf{P}}\tilde{\mathbf{P}}^H) = N_t. \end{aligned} \tag{11}$$

### A. Asymptotic Optimality and Concavity of Lower Bound

*1) Asymptotic Optimality at Low SNR Region:* When $\sigma^2 \rightarrow +\infty$, the objective function in (11) is expressed, based on Taylor expansion, as

$$\begin{aligned} &\sum_{m=1}^{M^{N_t}} \log \sum_{k=1}^{M^{N_t}} \prod_{q} \left(1 + \frac{r_q}{2\sigma^2} \mathbf{e}_{mk}^H \tilde{\mathbf{P}}^H \mathbf{\Sigma_t} \tilde{\mathbf{P}} \mathbf{e}_{mk}\right)^{-1} \\ &\quad = M^{N_t} \log M^{N_t} - \frac{\mathsf{Tr}(\mathbf{\Sigma_r})}{2\ln(2) M^{N_t} \sigma^2} \\ &\qquad \cdot \left( \sum_{m}^{M^{N_t}} \sum_{k}^{M^{N_t}} \mathbf{e}_{mk}^H \tilde{\mathbf{P}}^H \mathbf{\Sigma_t} \tilde{\mathbf{P}} \mathbf{e}_{mk} \right) + \mathcal{O}(1/\sigma^4) \end{aligned}$$

where $\mathcal{O}(1/\sigma^4)$ denotes the least-significant terms on the order of $1/\sigma^4$. Since $\mathbf{e}_{mk}^H \tilde{\mathbf{P}}^H \mathbf{\Sigma_t} \tilde{\mathbf{P}} \mathbf{e}_{mk}$ is a scalar, it yields

$$\sum_{m}^{M^{N_t}} \sum_{k}^{M^{N_t}} \mathbf{e}_{mk}^H \tilde{\mathbf{P}}^H \mathbf{\Sigma_t} \tilde{\mathbf{P}} \mathbf{e}_{mk} = e_{mk} \mathsf{Tr}\left(\mathbf{\Sigma_t} \mathbf{\Sigma_P^2}\right) \tag{12}$$

where $e_{mk}$ is a constant with $e_{mk}\mathbf{I} = \sum_{m}^{M^{N_t}} \sum_{k}^{M^{N_t}} \mathbf{e}_{mk}\mathbf{e}_{mk}^H$.

Combining the optimal precoding structure in Proposition 1 and the diagonal matrix $\mathbf{\Sigma_P}$ of maximizing (12) with power constraint, the solution of problem (11) at low SNR region is given by the following proposition:

*Proposition 3:* The optimal precoder to maximize the lower bound at low SNR region equals $\mathbf{U_t}$ times a diagonal power allocation matrix with all power allocated on the maximum singular value of $\mathbf{\Psi_t}$ (*i.e.*, the beamforming strategy).

The result based on maximizing the lower bound presented here is the same as the result of maximizing average mutual information directly at the low SNR region (the latter can be derived by extending the analysis for the case of instantaneous CSI in [4]); that is, it is asymptotically optimal to maximize the lower bound at the low SNR region.

*2) Asymptotic Optimality at High SNR Region:* The proof of asymptotic optimality at high SNR region starts by rewriting the objective function of (11) as

$$\sum_{m=1}^{M^{N_t}} \log \sum_{k=1}^{M^{N_t}} \exp\left[ -\sum_{q} \ln\left(1 + \frac{r_q}{2\sigma^2} \mathbf{e}_{mk}^H \tilde{\mathbf{P}}^H \mathbf{\Sigma_t} \tilde{\mathbf{P}} \mathbf{e}_{mk}\right)\right]. \tag{13}$$

Note that $\log \sum_{k=1}^{M^{N_t}} \exp(\cdot)$ is a *soft* version of maximization [11]. The idea here is to replace the *soft* maximization by its *hard* version and approximate it for high SNR region

$$\begin{aligned} &\sum_{m=1}^{M^{N_t}} \max_{k} \left[ -\sum_{q} \ln\left(1 + \frac{r_q}{2\sigma^2} \mathbf{e}_{mk}^H \tilde{\mathbf{P}}^H \mathbf{\Sigma_t} \tilde{\mathbf{P}} \mathbf{e}_{mk}\right)\right] \\ &\approx \sum_{m=1}^{M^{N_t}} \max_{k} \left[ -\sum_{q} \ln\left(\frac{r_q}{2\sigma^2}\right) - N_r \ln\left(\mathbf{e}_{mk}^H \tilde{\mathbf{P}}^H \mathbf{\Sigma_t} \tilde{\mathbf{P}} \mathbf{e}_{mk}\right)\right]. \end{aligned}$$

Thus, the problem in (11), at high SNR, is equivalent to

$$\begin{aligned} \text{maximize} \quad & \min_{\substack{m,k \\ m \neq k}} \mathbf{e}_{mk}^H \tilde{\mathbf{P}}^H \mathbf{\Sigma_t} \tilde{\mathbf{P}} \mathbf{e}_{mk} \\ \text{subject to} \quad & \mathsf{Tr}(\tilde{\mathbf{P}}\tilde{\mathbf{P}}^H) = N_t. \end{aligned} \tag{14}$$

The minimization of the quadratic form in (14) can be identified as the minimum distance among all possible realizations of the input vector

$$d_{\min} = \min_{\substack{m,k \\ m \neq k}} \|\mathbf{\Sigma_t}^{1/2} \tilde{\mathbf{P}} (\mathbf{x}_m - \mathbf{x}_k)\|^2 \tag{15}$$

which leads to the following results:

*Proposition 4:* The optimal precoder to maximize the lower bound at high SNR region is equivalent to maximizing the minimum distance among all the constellation vectors.

This result, maximizing the lower bound for high SNR, is the same as that of maximizing the average mutual information by extending [4]; that is, it is also asymptotically optimal to maximize the lower bound at the high SNR region.

*3) Concavity Results:* Considering the low computational complexity and the asymptotic optimality, it is reasonable to apply the criterion of maximizing the lower bound. In order to develop efficient algorithm, concavity, guaranteeing the global optimality, needs to be verified.

The first candidate is to identify the concavity of $\tilde{\mathbf{P}}$. Unfortunately, it does not hold and can be verified by a counterexample (*e.g.*, $\mathbf{H} \in \mathbb{C}^{1\times 1}$). The next candidate is to identify the concavity over $\mathbf{\Sigma_P^2}$.

*Proposition 5:* The lower bound of the average mutual information is a concave function of $\boldsymbol{\lambda}$, $\boldsymbol{\lambda} = \mathsf{Diag}(\mathbf{\Sigma_P^2})$.

*Proof:* This result can be proved by identifying the concavity over the diagonal elements of $\mathbf{\Sigma_P^2}$ in (13). □

### B. Precoder Design

The solution of $\tilde{\mathbf{P}}$ is separated into two parts: optimization of power allocation matrix and right singular vectors.

*Optimal Power Allocation*: Given right singular vectors $\mathbf{V_P}$, the first step is to optimize $\boldsymbol{\lambda}$:

$$\begin{aligned} \text{maximize} \quad & \mathcal{I}_L(\boldsymbol{\lambda}) \\ \text{subject to} \quad & \mathbf{1}^T \boldsymbol{\lambda} = N_t \\ & \boldsymbol{\lambda} \succeq \mathbf{0} \end{aligned} \tag{16}$$

where $\mathbf{1}$ and $\mathbf{0}$ denote the column vector with all entries one and zero, respectively. The concavity result in Proposition 5 ensures that a global optimal solution can be found by either gradient-descent based method or Newton-type method [11].

*Optimization Over Right Singular Vectors*: Given a power allocation vector $\boldsymbol{\lambda}$, the maximization of lower bound boils down to the maximization over the right singular vectors:

$$\begin{aligned} \text{maximize} \quad & \mathcal{I}_L(\mathbf{V_P}) \\ \text{subject to} \quad & \mathbf{V_P}^H \mathbf{V_P} = \mathbf{I}. \end{aligned} \tag{17}$$

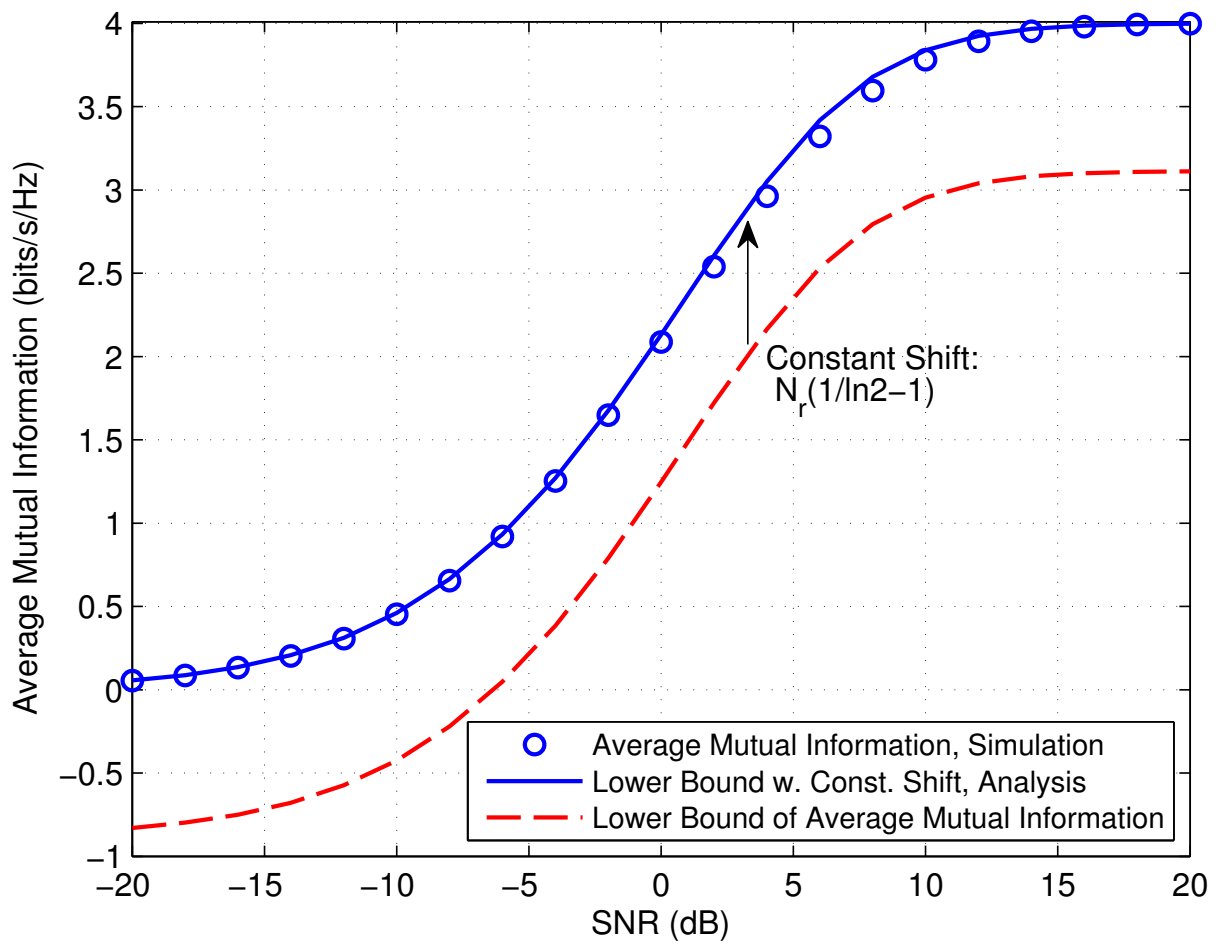

Fig. 1. Average mutual information with QPSK inputs and exponentially correlated ($\rho_t = 0.8$, $\rho_r = 0.5$) MIMO channels ($N_t = N_r = 2$) for the case of without precoding.

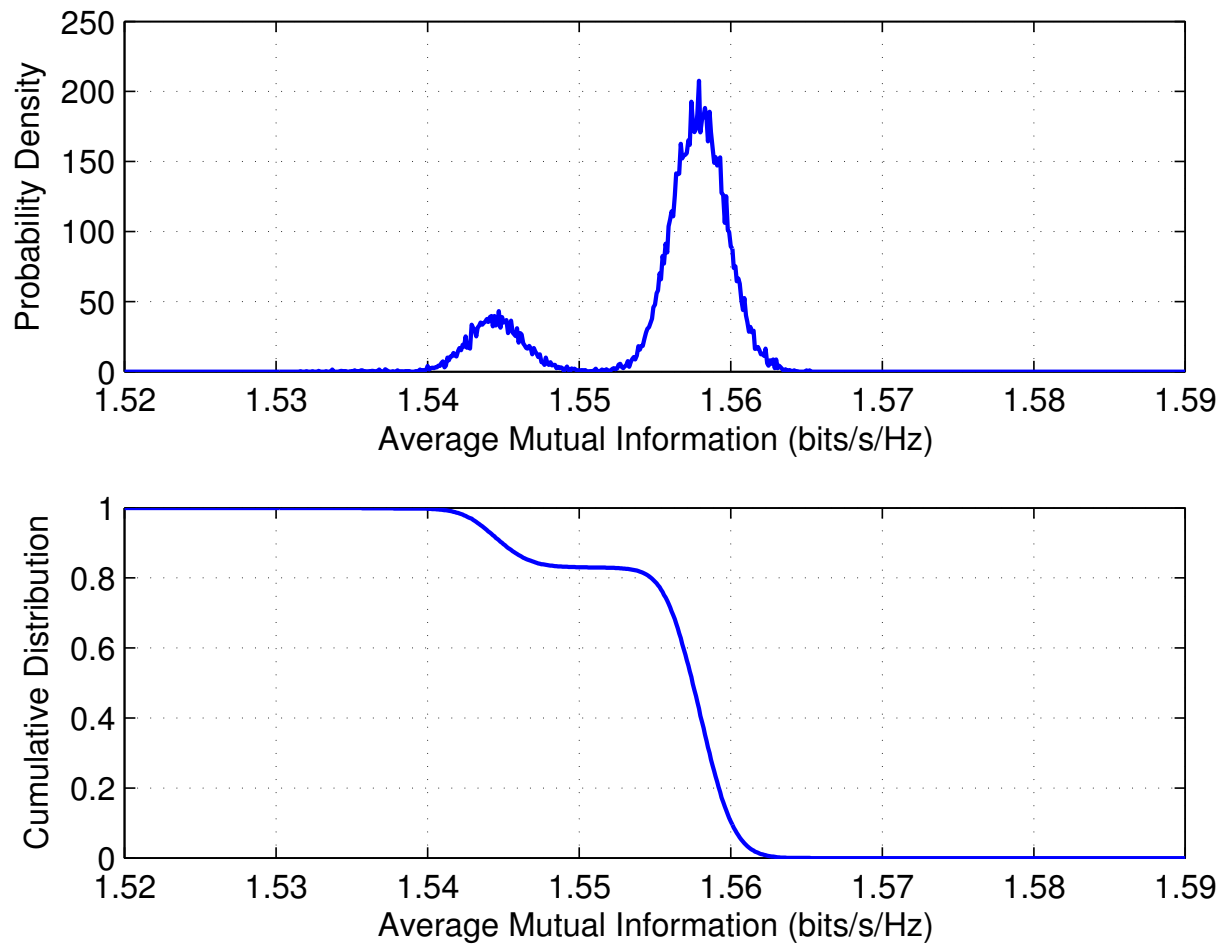

Fig. 2. Probability density and cumulative distribution of the optimized average mutual information from different initialization points. The input signal is drawn from QPSK; MIMO channels ($2 \times 2$) are correlated with $\rho_t = 0.8$ and $\rho_r = 0.5$; SNR is -5 dB.

To solve the above unitary-matrix constrained problem, both gradient descent with projection and moving towards geodesics on Riemannian manifolds can be used [12].

*Two-Step Approach to Optimize Precoder*: A two-step approach can now be developed by combining the design for left singular vectors in Proposition 1, power allocation vector in problem (16), and right singular vectors in problem (17).

*Algorithm:* Two-Step Algorithm to Maximize the Lower Bound of Average Mutual Information Over Linear Precoder

1) **Initialization**. Given feasible initial points $\boldsymbol{\lambda}^{(0)}$ and $\mathbf{V}_{\mathbf{P}}^{(0)}$, and set $n = 1$.
2) **Design left singular vectors**. Let $\mathbf{U_P} = \mathbf{U_t}$.
3) **Update power allocation matrix**. Solve problem (16):
$$\boldsymbol{\lambda}^{(n)} = \arg \max_{\substack{\mathbf{1}^T\boldsymbol{\lambda}=N_t \\ \boldsymbol{\lambda}\succeq 0}} \mathcal{I}_L(\boldsymbol{\lambda}, \mathbf{V}_{\mathbf{P}}^{(n-1)}).$$
4) **Update right singular vectors**. Solve problem (17):
$$\mathbf{V}_{\mathbf{P}}^{(n)} = \arg \max_{\mathbf{V}_{\mathbf{P}}^H\mathbf{V}_{\mathbf{P}}=\mathbf{I}} \mathcal{I}_L(\boldsymbol{\lambda}^{(n)}, \mathbf{V_P}).$$
5) **Iteration**. Set $n = n + 1$, and go to Step 3 until convergence.

The two-step algorithm, optimizing variables alternatively, converges to the globally optimum solution when the optimal right singular vectors are unique and the bound is concave on $\mathbf{V_P}$. When this condition fails, the iterative algorithm converges to a local maximum, which may be affected by the initialization of the algorithm. However, we show, by numerical examples in the next section, that the different initializations have limited effect on solution; that is, the two-step algorithm achieves near-optimal performance.

## V. Simulation Results

Examples are provided to illustrate the relationship between average mutual information and the derived bound and to show the convergence and the efficacy of the proposed algorithm. To exemplify our results, the exponential correlation model
$$[\boldsymbol{\Psi}(\rho)]_{i,j} = \rho^{|i-j|}, \quad \rho \in [0, 1)$$
with $\boldsymbol{\Psi}_\mathbf{t} = \boldsymbol{\Psi}(\rho_t)$ and $\boldsymbol{\Psi}_\mathbf{r} = \boldsymbol{\Psi}(\rho_r)$ is considered.

*1) Relationship between Average Mutual Information and Lower Bound:* When the SNR approaches $0$ and $+\infty$, the limits of the average mutual information in (3) are given by $0$ and $N_t \log M$. At the same time, the limits of the lower bound in Proposition 2 are, respectively, $-N_r(1/\ln(2) - 1)$ and $N_t \log M - N_r(1/\ln(2) - 1)$, which imply a constant gap at low and high SNR regions between average mutual information and the lower bound. Since adding a constant value to the lower bound function remains the solution to the optimization problem (11) invariant, with a constant shift, the lower bound actually serves as a very good approximation. Figure 1 illustrates the derived lower bound, lower bound with a constant shift, $N_r(1/\ln(2) - 1)$, and the simulated average mutual information (by the Monte Carlo method via generating many realizations of $\mathbf{H_w}$ and $\mathbf{n}$, see (3) for formula). The lower bound with a shift and the simulated curve match exactly at low and high SNR regions and close to each other at medium SNR region. Further study verifies that this approximation is valid for various numbers of transmit and receive antennas and various input types and correlation parameters [10]. These results imply the precoder maximizing the lower bound can be a good solution.

*2) Convergence of the Two-Step Algorithm:* The convergence of the algorithm is considered with different feasible initialization points. The initial power allocation vector $\boldsymbol{\lambda}^{(0)}$ is non-negative and satisfies sum power constraint, while the initial right singular vectors satisfy unitary constraint.

As the lower bound is optimized iteratively by the proposed algorithm, the average mutual information is forced to improve. The probability density and cumulative distribution of average mutual information for the optimized linear precoder from different initialization points are depicted in Fig. 2, which is obtained by generating 300,000 uniform random initial power allocation matrices and right singular vectors. The curve of probability density implies the existence of multiple local optimum points, which verify the nonconcavity result over $\mathbf{P}$

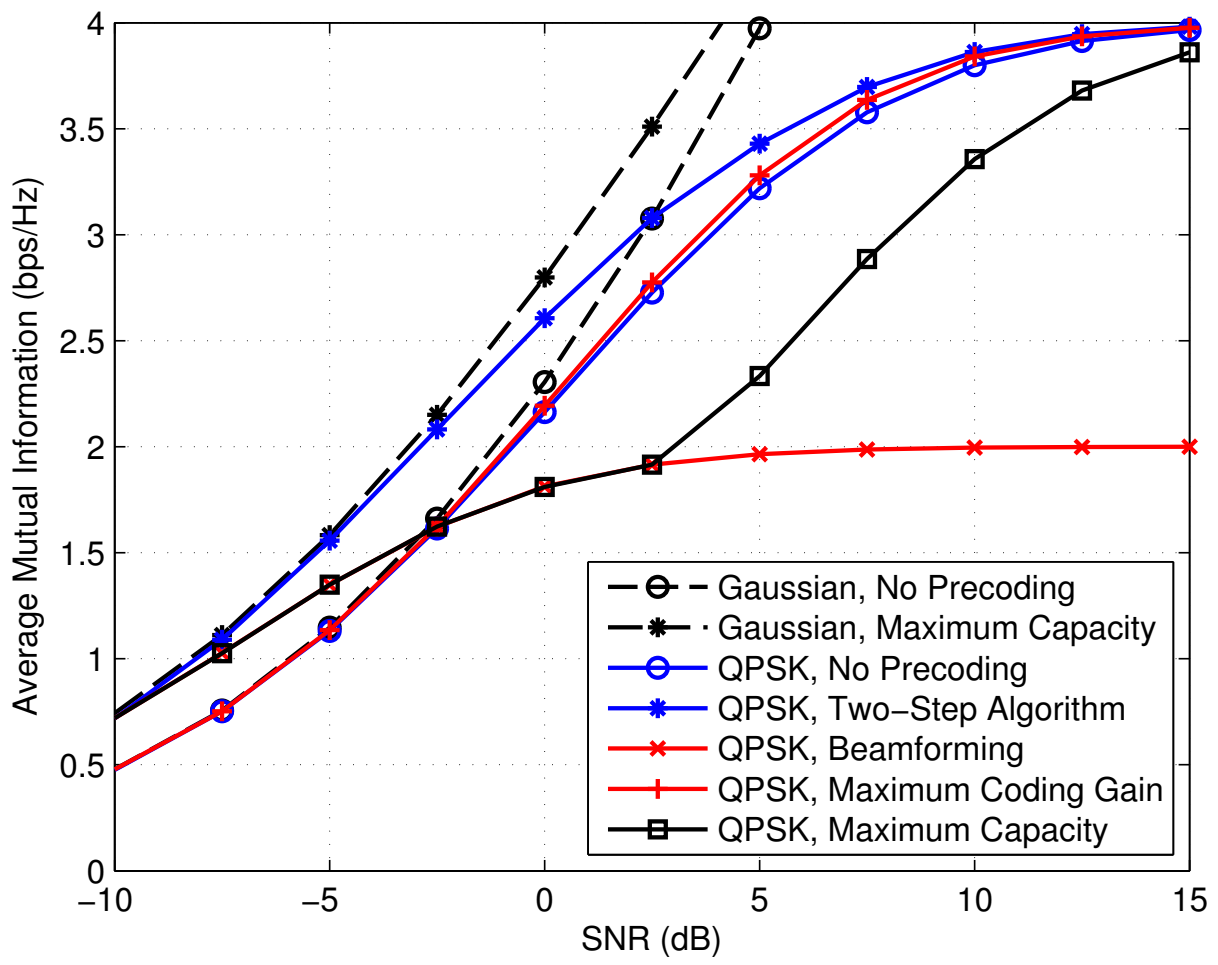

Fig. 3. Average mutual information versus the SNR for different strategies. The input signal is drawn from QPSK; $2\times2$ MIMO channels are exponentially correlated with $\rho_t = 0.8$ and $\rho_r = 0.5$.

(see Sec. IV-A3). Although there is a likelihood to stop at a local optimum, the two-step algorithm, from an arbitrary initialization points, achieves average mutual information more than 1.535 bps/Hz, about 97% of the maximum capacity with Gaussian input [13] (1.583 bps/Hz, see Fig. 3 for reference). That is, the iterative algorithm obtains a satisfactory solution, even though the problem is nonconcave, and makes performance of MIMO systems with finite-alphabet inputs close to the maximum capacity with Gaussian inputs.

*3) Efficacy of the Linear Precoder:* The performance of the proposed algorithm is shown in Fig. 3, which also includes several additional cases: no precoding for both QPSK inputs and Gaussian inputs, beamforming, maximum capacity [13], and maximum coding gain [14].

Although the precoding method of maximum capacity obtains gains when input signal is from Gaussian distribution, it results in a significant *loss* if applying the strategy to discrete inputs, especially at the medium to high SNR region. The reason for such performance comes from differences in designing the power allocation matrix and right singular vectors between finite-alphabet inputs and Gaussian inputs.

Intuitively, in order to maximize capacity with Gaussian inputs, allocating more power to the stronger subchannels and less to the weaker subchannels is the solution. This design, however, is not optimal for finite-alphabet inputs because the average mutual information with finite inputs is bounded, and allocating more power to subchannels that close to saturation is not efficient. Moreover, the right singular vectors for Gaussian inputs is an arbitrary unitary matrix, while the case of finite inputs fails to follow the same rule, as shown in (4).

The proposed precoding algorithm exploits the characterization of the optimal precoding structure and achieves a solution with the optimal left singular vectors, the optimal power allocation vector (given an arbitrary right singular vectors), and a local optimal right singular vectors. Since different initialization points have limited effect on solution (see Fig. 2), the two-step algorithm guarantees to offer a significant gain from an arbitrary initialization point. For example, performance of the proposed algorithm is about 1.9 dB, 2.2 dB, and 5.9 dB better than the maximum coding gain, no precoding, and maximum capacity method, respectively, when channel coding rate is 1/2. Moreover, when SNR is less than -2.5 dB, the proposed algorithm provides almost the same performance as the maximum capacity design with Gaussian inputs, which is the upper bound for all possible linear precoder.

## VI. Conclusion

This paper has considered the linear precoding over MIMO channels with statistical CSI. Instead of assuming Gaussian inputs, theoretically optimal but rarely realized in practice, it has explored the framework to maximize the average mutual information with the constraint of finite-alphabet inputs, which has been known as an important open problem. The obstacles includes two aspects: First, the closed form of average mutual informations is lacking; second, the optimization problem over precoding matrix is nonconcave. Both obstacles have made the problem of finding a good solution extremely difficult. This study has exploited the structure of the optimal precoder and solved this problem by a unified two-step iterative algorithm. Numerical examples have demonstrated the convergence and performance of the proposed algorithm. Compared to its conventional counterparts, the linear precoding method can provide a significant performance gain.